\documentclass[12pt]{iopart}

\usepackage{graphicx}
\usepackage{amsfonts}
\usepackage{epsfig}
\usepackage{graphicx}% Include figure files
\usepackage{dcolumn}% Align table columns on decimal point
\usepackage{bm}% bold math
\usepackage{times}
\usepackage{xcolor}

\begin{document}

\title{Optimal community structure for social contagions}

\author{Zhen Su$^{1,2}$, Wei Wang$^{3}$, Lixiang Li$^{4}$, H. Eugene
  Stanley$^{5}$, Lidia A. Braunstein$^{5,6}$} 

\address{$^{1}$ College of Computer Science and Technology, Chongqing
  University of Posts and Telecommunications, Chongqing 400065, China}

\address{$^{2}$ Chongqing MII Key Lab. of Computer Networks \&
  Communications, Chongqing 400065, China} 

\address{$^{3}$ Cybersecurity Research Institute, Sichuan University,
  Chengdu 610065, China} 

\address{$^{4}$ Information Security Center, State Key Laboratory of
  Networking and Switching Technology, Beijing University of Posts and
  Telecommunications, Beijing 100876, China}

\address{$^{5}$ Center for Polymer Studies and Department of Physics,
Boston University, Boston, Massachusetts 02215, USA}

\address{$^{6}$ Instituto de Investigaciones F\'isicas de Mar del
  Plata (IFIMAR)-Departamento de F\'isica, Facultad de Ciencias
  Exactas y Naturales, Universidad Nacional de Mar del Plata-CONICET,
  Funes 3350, (7600) Mar del Plata, Argentina}

\address{E-mail: wwzqbx@hotmail.com, li\_lixiang2006@163.com}

%\ead{custserv@iop.org}

\begin{abstract}

\noindent
Community structure is an important factor in the behavior of real-world
networks because it strongly affects the stability and thus the phase
transition order of the spreading dynamics. We here propose a reversible
social contagion model of community networks that includes the factor of
social reinforcement. In our model an individual adopts a social
contagion when the number of received units of information exceeds its
adoption threshold. We use mean-field approximation to describe our
proposed model, and the results agree with numerical simulations.  The
numerical simulations and theoretical analyses both indicate that there
is a first-order phase transition in the spreading dynamics, and that a
hysteresis loop emerges in the system when there is a variety of
initially-adopted seeds. We find an optimal community structure that
maximizes spreading dynamics. We also find a rich phase diagram with a
triple point that separates the no-diffusion phase from the two
diffusion phases.

\end{abstract}


\maketitle

\tableofcontents

\section{Introduction}

\noindent
Social contagion---including the spreading of social information,
opinions, cultural practices, and behavior patterns---is ubiquitous in
nature and society \cite{cas,Boccaletti, Guardiola, Ohta}.  Unlike
biological contagion \cite{psr,daqing}, social reinforcement, which is
also ubiquitous, plays a central role in social contagions and triggers
such complex dynamic phenomena \cite{wang2017, pastor2015,su} as
first-order phase transitions \cite{gao2012}. Empirical studies indicate
that susceptible individuals adopt a social behavior only when the
number of received information units exceeds an adoption threshold
\cite{watts2002,Centola,Banerjee,lee}. Thus this behavior occurs when a
certain level of exposure is exceeded. The numerous Markovian and
non-Markovian models of complex networks used to describe social
contagion \cite{wang2016,ps2015,Castellano09,Dorogovtsev08} indicate
that the topology of networks strongly affects patterns of social
contagion
\cite{Gleeson2007,Whitney,Gleeson2008,Nematzadeh,Lee,Brummitt,wangnjp2,Boccaletti1,wangzhen}.
Recently scholars extended the social contagion model to multiplex
networks and found that multiplexity promotes social contagion
\cite{Lee2014,Brummitt2012,yagan2012}. Holme et al.
\cite{Takaguchi,Karimi} found that a temporal network in which the
network structure changes with time can either promote or suppress
social contagions under various scenarios.  Macroscopically, researchers
have found that the average degree and the level of heterogeneity of the
degree distribution changes the growth patterns of social contagions
\cite{wangpre,wangnjp}.  Microscopically, social contagions exist in a
hierarchy \cite{wangpre}, i.e., high-degree nodes or hubs are infected
in the early stages of the infection process and low-degree nodes in the
later stages. Mesoscopically, researchers have studied how degree
correlation and community structure affect social contagion
\cite{Radicchi,Doddspre}. Researchers have found a level of network
modularity---the measurement of how strongly a network is divided into
modules or communities---that is optimal. The initial number of adopter
seeds that allows a global diffusion of the contagion is at its minimum
\cite{nem}. Majdandzic proposed a contagion model with an adoption
threshold and spontaneous adoption, and found the system has hysteresis
loop and phase-flipping \cite{Antonio}.

Most previous studies have focused on an irreversible social contagion
in which infected agents either recover or die and in both cases no
longer can be infected \cite{peters,psa}. These studies do not take into
account the effect of reversible social contagion in which infected
agents can once again be infected after passing through a susceptible
period \cite{liu}. In real-world epidemics \cite{rps2001} individuals
often are not fully immunized and return to a susceptible state after
having been infected. We here present a reversible social contagion
model of a community network \cite{newmans1,newmans2}. Initially a
number of infected individuals are randomly distributed in the
community. All other individuals are susceptible. Susceptible
individuals become infected when the number of received information
units exceeds their adoption thresholds. We derive our model using
mean-field theory.  Both numerical simulations and theoretical analyses
indicate the presence of a hysteresis loop in social contagions. More
important, we find an optimal network modularity that globally promotes
social contagions. The constant threshold point, the critical threshold
fraction of intracommunity links, triggers a sharp transition from a
no-diffusion state to a global diffusion state.

This paper is organized as follows. In Sec.~2, we propose a social
contagion model for community networks. In Sec.~3 we develop a
mean-field theory to mathematically analyze our model.  In Sec.~4 we
simulate the proposed model on a community network and show the results.
In Sec.~5 we discuss our conclusions.

\section{Model descriptions} \label{Model}

\noindent
In our model the network has two equal-sized communities, $a$ and $b$,
with $N$ nodes and $L$ links in the network system. Initially nodes are
with equal probability assigned to either community $a$ or community
$b$. Then $(1-\mu)L$ links are randomly distributed among node pairs
within a community and $\mu L$ are randomly distributed among node pairs
between communities $a$ and $b$. The $\mu$ value is the probability that
a randomly selected link is an interlink between different
communities. We adjust the strength of the social community by changing
the value of $\mu$. Figure~1(d) shows a matrix of the community. Matrix
$\mathcal{A}$ ($\mathcal{D}$) shows the connections among individuals
within community $a$ ($b$). Matrix $\mathcal{C}$ ($\mathcal{B}$) shows
the individuals in community $b$ ($a$) connected to individuals in
community $a$ ($b$).

Using this topology we develop a susceptible-adopted-susceptible (SAS)
social contagion model of a community network. Individuals are either
susceptible (S) or adopted (A). A susceptible individual can receive
information from adopted neighbors in communities $a$ and $b$. An
adopted individual can transmit the social contagion to susceptible
neighbors. At the initial stage, a random fraction of $\rho_0$ of
individuals are adopted in community $a$, and the remaining individuals
are susceptible in both communities. An adopted individual has adopted
the behavior and with probability $\lambda$ transmits the information to
susceptible neighbors that belong to both communities. If the units of
information $m$ a susceptible individual has received exceeds an
adoption threshold $\theta$, the susceptible individual enters the
adopted state. The parameter $\theta$ indicates the willingness of an
individual to adopt a new behavior. Large (small) $\theta$ values
indicate that susceptible individuals need a large (small) amount of
information before they enter into the adopted state. Each adopted
individual with probability $\gamma$ loses interest in the social
contagion and returns to the susceptible state. Figures~1(a)--1(c)
schematically show this information spreading process.

\begin{figure}
\begin{center}
  \includegraphics[scale=0.4]{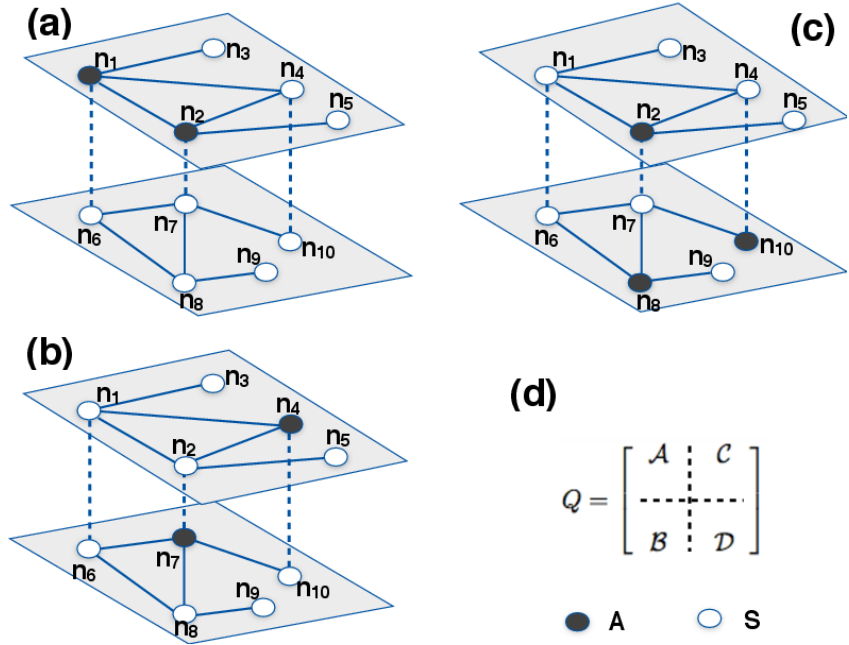}
\caption{(Color online) Schematic of two-community system where the
  contagion dynamics take place. The agents $n_1$, $n_2$, $n_3$, $n_4$
  and $n_5$ are in community $a$, and $n_6$, $n_7$, $n_8$, $n_9$ and
  $n_{10}$ are in community $b$. (a) At time step $1$ , the agents $n_1$
  and $n_2$ are in the adopted state, and the other agents are in the
  susceptible state. (b) At time step $2$, susceptible agents $n_4$ and
  $n_7$ change to the adopted state because the number of received
  information $m$ exceed the adoption threshold $\theta$. At the same
  time, the adopted agents $n_1$ and $n_2$ recover and go to the
  susceptible state. (c) At time step $3$, the susceptible agents $n_2$,
  $n_8$ and $n_{10}$ enter the adopted state because the number of
  received information $m$ exceed the adoption threshold $\theta$. At
  the same time, the adopted agents $n_4$ and $n_7$ recover and go to
  the susceptible state. (d) $Q$ is the adjoin matrix of the
  system. $\mathcal{A}$, $\mathcal{B}$, $\mathcal{C}$ and $\mathcal{D}$
  are the partitioned matrix of $Q$.
\label{fig1}}
\end{center}
\end{figure}

\section{Theory} \label{theory}

\subsection{Mathematical theory}

\noindent
Here we derive a mean-field theory for our model that reproduces social
contagion dynamics. We denote $\rho^{\ell}_i(t)$ ($\ell=a$ or $b$) to be
the density of individuals in community $\ell$ in the adopted state at
time $t$. The dynamic equations for $\rho^{a}_i(t)$ and $\rho^{b}_i(t)$
are
\begin{equation}\label{T1}
\frac{d\rho^{a}_i(t)}{dt}=-\gamma\rho^{a}_i(t)+[1-
  \rho^{a}_i(t)]\delta(\lambda \sum_j \mathcal{A}_{ij} \rho_j^{a}(t)
+\lambda  \sum_j \mathcal{C}_{ij} \rho_j^{b}(t)),
\end{equation}
and
\begin{equation}\label{T2}
\frac{d\rho^{b}_i(t)}{dt}=-\gamma\rho^{b}_i(t)+[1-
  \rho^{b}_i(t)]\delta(\lambda \sum_j \mathcal{B}_{ij} \rho_j^{a}(t) +
\lambda \sum_j \mathcal{D}_{ij} \rho_j^{b}(t)),
\end{equation}
respectively. Here $\gamma\rho^{\ell}_i(t)$ is the probability that an
adopted individual $i$ recovers at time $t$ in community $\ell$, and
$\lambda \sum_j \mathcal{A}_{ij} \rho_j^{a}$ and $\lambda \sum_j
\mathcal{C}_{ij} \rho_j^{b}$ respectively are the units of information a
susceptible individual $i$ in community $a$ receives from adopted
neighbors in communities $a$ and $b$ at time $t$. We set $\lambda \sum_j
\mathcal{B}_{ij} \rho_j^{a}$ and $\lambda \sum_j \mathcal{D}_{ij}
\rho_j^{b}$ to respectively represent the units of information a
susceptible individual $i$ in community $b$ receives from adopted
neighbors in communities $a$ and $b$ at time $t$. The function $\delta$
is the probability that an individual becomes adopted. Thus
$\delta(m)=1$ when the information received by an individual ($m$)
exceeds the adoption threshold ($\theta$), i.e., when $m \ge \theta$ and
zero otherwise.

Using Eqs.~(\ref{T1}) and (\ref{T2}) we determine the evolution of
social contagions in community networks.  Note that we need $N$
differential equations to describe the spreading dynamics. When
$N\rightarrow \infty$, it is difficult to solve the equations. More
important, it is difficult to determine the transition points of the
system. For simplicity we assume $\rho_i^{\ell} \equiv
\rho_i^{\ell}(t)$, $F(\rho_i^{\ell})=-\gamma \rho_i^{\ell}(t)$, and
$K(\rho_i^{\ell})=[1- \rho_i^{\ell}(t)]$. Equations~(1) and (2) can be
written in terms of $F(\rho_i^{\ell})$ and $K(\rho_i^{\ell})$ as
\begin{equation}\label{rho}
\frac{d\rho^{a}_i}{dt}=F(\rho_i^a)+K(\rho_i^a)\delta(\lambda \sum_{j}
\mathcal{A}_{ij}\rho_j^a+\lambda \sum_{j} \mathcal{C}_{ij}\rho_j^b),
\end{equation}
and
\begin{equation}\label{rho0}
\frac{d\rho^{b}_i}{dt}=F(\rho_i^b)+K(\rho_i^b)\delta(\lambda \sum_{j}
\mathcal{B}_{ij}\rho_j^a+\lambda \sum_{j} \mathcal{D}_{ij}\rho_j^b).
\end{equation}
These equations describe the dynamic interactions of all nodes in the
system. Calculating the time-dependent activities of all the interactive
nodes is complex. A susceptible high-degree individual $i$ is more
likely to receive information from neighbors than a susceptible
small-degree individual $j$. Thus the probability that susceptible
individual $i$ receives information from neighbor $j$ is proportional to
the degree of $j$. Using Ref.~\cite{gao2016} we evaluate the dynamic
evolution process of a node by quantifying the average dynamics of
neighbor nodes. The degree of node $j$ is $s_j^Q=\sum_{i=1}^N Q_{ij}$
($Q$ is the adjacency matrix of the system). We introduce $\langle y_j
\rangle_{nn}$ with the scalar quantity $y_j$ related to the degree of
node $j$
\begin{eqnarray} \label{y2}
 \langle y_j \rangle _{nn}
&=\frac{\frac{1}{N} \sum_{j=1}^{N} s_{j}^My_j}
{\frac{1}{N} \sum_{j=1}^{N}s_{j}^M}=\frac{\bm{I^TMy}}{\bm{I^TMI}} \nonumber\\
&=\frac{ \langle s_j^M y_j\rangle}{ \langle s_j^M \rangle}=\mathfrak{C}(\bm{y})_M,
\end{eqnarray}
where $M\in\{\mathcal{A},\mathcal{B},\mathcal{C},\mathcal{D}\}$,
$\bm{I}=(1,\ldots,1)^T$, $\bm{y}=(y_1,\ldots,y_N)^T$, and
$\mathfrak{C}(\bm{y})_M$ is an operator, which is the nearest neighbor
average to the explicit summation. From Eq.~(\ref{y2}) we know that
higher degree nodes contribute more to $\langle y_j \rangle _{nn} $. If
we assume $y_j(\rho_j^{\ell}) = \rho_j^{\ell}$,
Eqs.~(\ref{rho}) and (\ref{rho0}) can be rewritten
\begin{equation}\label{rho3}
\frac{d\rho^{a}_i}{dt}
=F(\rho_i^a)+K(\rho_i^a)\delta(\lambda s_{i}^{\mathcal{A}}
\mathfrak{C}(\bm{\rho_a})_{\mathcal{A}}+\lambda s_{i}^{\mathcal{C}}
\mathfrak{C}(\bm{\rho_b})_{\mathcal{C}}),
\end{equation}
and
\begin{equation}\label{rho4}
\frac{d\rho^{b}_i}{dt}
=F(\rho_i^b)+K(\rho_i^b)\delta(\lambda s_{i}^{\mathcal{B}}
\mathfrak{C}(\bm{\rho_a})_{\mathcal{B}}+\lambda s_{i}^{\mathcal{D}}
\mathfrak{C}(\bm{\rho_b})_{\mathcal{D}}),
\end{equation}
where $\bm{\rho_{\ell}}=(\rho_1^{\ell}, \rho_2^{\ell}, \cdots,
\rho_n^{\ell})^{\rm T}$. Inspired by Ref.~\cite{gao2016} we use
equations Eqs.~(\ref{rho3}) and (\ref{rho4}) to describe the spreading
dynamics and rewrite them in terms of vectors,
\begin{equation}\label{rho5}
\frac{d\bm{\rho_{a}}}{dt}=F(\bm{\rho_a})+K(\bm{\rho_a})\delta(\lambda
\bm{s_{\mathcal{A}}}  \mathfrak{C}(\bm{\rho_a})_\mathcal{A} + \lambda
\bm{s_{\mathcal{C}}}  \mathfrak{C}(\bm{\rho_b})_\mathcal{C} ),
\end{equation}
and
\begin{equation}\label{rho6}
\frac{d\bm{\rho_{b}}}{dt}=F(\bm{\rho_b})+K(\bm{\rho_b})\delta(\lambda
\bm{s_{\mathcal{B}}} \mathfrak{C}(\bm{\rho_a})_\mathcal{B} + \lambda
\bm{s_{\mathcal{D}}} \mathfrak{C}(\bm{\rho_b} )_\mathcal{D}),
\end{equation}
where $\bm{s_{M}}=(s_1^{M}, s_2^{M}, \cdots, s_n^{M})^{\rm T}$.  From
Eqs.~(\ref{rho5}) and (\ref{rho6}) we obtain the fraction of infected
nodes. When $t\rightarrow\infty$ we denote the final behavior adoption
size in community $a$ and $b$ to be $\rho_a$ and $\rho_b$,
respectively. The final behavior adoption size of the system is
$\rho=\rho_a+\rho_b$.

\subsection{Threshold points}

\noindent
Another important factor in the spreading dynamics concerns any existing
threshold points. To obtain them we linearize Eqs.~(\ref{rho5}) and
(\ref{rho6}) around $\bm\rho_\ell=0$ ($\ell\in\{a,b\}$),
\begin{eqnarray}\label{rho7}
\frac{d\mathfrak{C}(\bm{\rho_a})_\mathcal{M}}{dt} &=
F(\mathfrak{C}(\bm{\rho_a})_\mathcal{M}) + K(\mathfrak{C}(\bm{\rho_a})_\mathcal{M})
 \delta(\lambda
 \mathfrak{C}(\bm{s_{\mathcal{A}}})_\mathcal{M}\mathfrak{C}(\bm{\rho_a})_\mathcal{A}
 \nonumber\\
&+\lambda \mathfrak{C}(\bm{s_{\mathcal{C}}})_{\mathcal{M}}
 \mathfrak{C}(\bm{\rho_b})_{\mathcal{C}}),
\end{eqnarray}
and
\begin{eqnarray}\label{rho8}
\frac{d\mathfrak{C}(\bm{\rho_b})_{\mathcal{N}}}{dt} &=
F(\mathfrak{C}(\bm{\rho_b})_{\mathcal{N}}) +
K(\mathfrak{C}(\bm{\rho_b})_{\mathcal{N}})  \delta(\lambda \mathfrak{C}
(\bm{s_{\mathcal{B}}})_{\mathcal{N}}
\mathfrak{C}(\bm{\rho_a})_{\mathcal{B}}\nonumber\\
 &+ \lambda \mathfrak{C}(\bm{s_{\mathcal{D}}})_{\mathcal{N}}
\mathfrak{C}(\bm{\rho_b})_{\mathcal{D}}),
\end{eqnarray}
where $\mathcal{M}\in\{\mathcal{A},\mathcal{B}\}$, and
$\mathcal{N}\in\{\mathcal{C},\mathcal{D}\}$. To obtain the threshold
points, we solve the above system with $N$ equations, but it is
difficult to obtain the analytic value. Thus we reduce the
dimensionality of the system by introducing an operator \cite{gao2016}.

The probability $\rho_{{\rm eff},M}^{\ell}$ that nodes in community
$\ell$ are infected by neighbors in community
$M\in\{\mathcal{A},\mathcal{B},\mathcal{C}, \mathcal{D}\}$ is
\begin{equation}\label{rho9}
 \rho_{{\rm
     eff},M}^{\ell}=\mathfrak{C}(\bm{\rho_{\ell}})_{M}=\frac{\bm{I^T M
     \rho_{\ell}}}{\bm{I^T M I}}=\frac{\langle s^{M}_j \rho^{\ell}_j
   \rangle}{\langle s^{M}_j \rangle}.
\end{equation}
We define $\beta_{U,M}$ ($U\in\{\mathcal{A}, \mathcal{B}, \mathcal{C}$,
$\mathcal{D}\}$) to be
\begin{equation}\label{rho10}
\beta_{U,M}=\mathfrak{C}(\bm{s_{U}})_{M}=\frac{\bm{I^T M
    s_{U}}}{\bm{I^TMI}}=\frac{\langle s^{M}_j s^{U}_j \rangle}{\langle
  s^{M}_j \rangle}.
\end{equation}
Inserting Eqs.~(\ref{rho9}) and (\ref{rho10}) into Eqs.~(\ref{rho7}) and
(\ref{rho8}), we obtain
\begin{equation}\label{rho11}
\frac{{d\rho_{{\rm eff},\mathcal{M}}^a}}{dt}=F(\rho_{{\rm
    eff},\mathcal{M}}^a)+K (\rho^a_{{\rm
    eff},\mathcal{M}})\delta(\lambda \beta_{\mathcal{A},\mathcal{M}}
\rho_{{\rm eff},\mathcal{A}}^a + \lambda \beta_{\mathcal{C},\mathcal{M}}
\rho_{{\rm eff},\mathcal{C}}^b),
\end{equation}
and
\begin{equation}\label{rho12}
\frac{{d\rho_{{\rm eff},\mathcal{N}}^b}}{dt}=F(\rho_{{\rm
    eff},\mathcal{N}}^b)+K (\rho^b_{{\rm eff},\mathcal{N}})
\delta(\lambda \beta_{\mathcal{B},\mathcal{N}} \rho_{{\rm
    eff},\mathcal{B}}^a + \lambda \beta_{\mathcal{D},\mathcal{N}}
\rho_{{\rm eff},\mathcal{D}}^b). \\
\end{equation}
In the steady state we have $d\rho_{{\rm eff},\mathcal{M}}^a/dt=0$ and
$d\rho_{{\rm eff},\mathcal{N}}^b/dt=0$. Thus we have
\begin{equation}\label{rho11}
f(\rho_{{\rm eff},\mathcal{M}}^a,\rho_{{\rm
    eff},\mathcal{N}}^b)=F(\rho_{{\rm eff},\mathcal{M}}^a)+K
(\rho^a_{{\rm eff},\mathcal{M}})\delta(\lambda
\beta_{\mathcal{A},\mathcal{M}} \rho_{{\rm eff},\mathcal{A}}^a + \lambda
\beta_{\mathcal{C},\mathcal{M}} \rho_{{\rm eff},\mathcal{C}}^b),
\end{equation}
and
\begin{equation}\label{rho12}
g(\rho_{{\rm eff},\mathcal{M}}^a,\rho_{{\rm
    eff},\mathcal{N}}^b)=F(\rho_{{\rm eff},\mathcal{N}}^b)+K
(\rho^b_{{\rm eff},\mathcal{N}}) \delta(\lambda
\beta_{\mathcal{B},\mathcal{N}} \rho_{{\rm eff},\mathcal{B}}^a + \lambda
\beta_{\mathcal{D},\mathcal{N}} \rho_{{\rm eff},\mathcal{D}}^b).  \\
\end{equation}
The Jacobian matrix of Eqs.~(\ref{rho11}) and (\ref{rho12}) is
\begin{equation} \label{Jacobian}
J=
\left(
 \begin{array}{cc}
 \frac{\partial f(\rho_{{\rm eff},\mathcal{M}}^a,
 \rho_{{\rm eff},\mathcal{N}}^b)}{\partial
 \rho_{{\rm eff},\mathcal{M}}^a}&
 \frac{\partial f(\rho_{{\rm eff},\mathcal{M}}^a,
 \rho_{{\rm eff},\mathcal{N}}^b)}{\partial
 \rho_{{\rm eff},\mathcal{N}}^b}\\
 \frac{\partial g(\rho_{{\rm eff},\mathcal{M}}^a,
 \rho_{{\rm eff},\mathcal{N}}^b)}{\partial
 \rho_{{\rm eff},\mathcal{M}}^a}&
 \frac{\partial g(\rho_{{\rm eff},\mathcal{M}}^a,
 \rho_{{\rm eff},\mathcal{N}}^b)}{\partial
 \rho_{{\rm eff},\mathcal{N}}^b}
\end{array}
\right).
\end{equation}
If adopted individuals have thresholds with $\lambda$, the determinant
of matrix $J$ equals zero. From Eq.~(\ref{Jacobian}) we obtain the
threshold information transmission probability $\lambda_c^{\rm inv}$ and
$\lambda_c^{\rm pre}$.

\section{Numerical verification}

\noindent
In this section we perform extensive simulations of an artificial
community network. We set the network size $N = 10^6$, the average
degree of each community $\left\langle k \right\rangle =20$, the
recovery probability $\gamma = 0.1$, and the adoption threshold
$\theta=5$. The initially adopted seeds $\rho_0$ are only in community
$a$.

\begin{figure}
\begin{center}
 \includegraphics[scale=0.6]{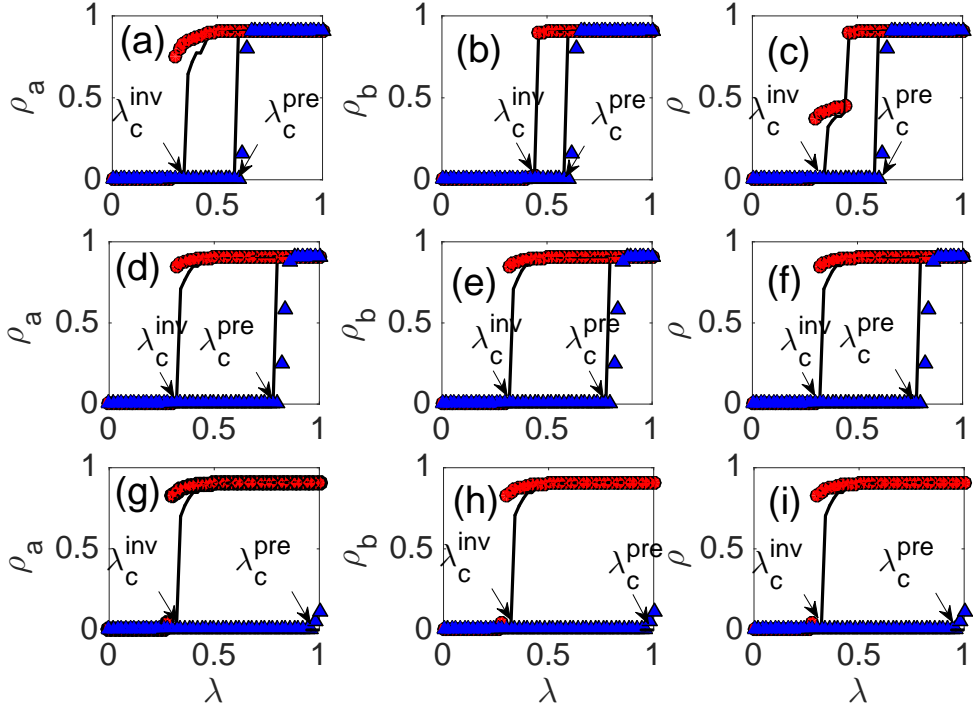}
 \caption{(Color online) The final behavior adoption size versus
   different strength of community structures. The final behavior
   adoption sizes with $\mu=0.1$ in Figs.~(a)--(c), $\mu=0.3$ in
   Figs.~(d)--(f), and $\mu=0.5$ in Figs.~(g)--(i).  The lines are the
   theoretical predictions.  The arrows represent the presence threshold
   $\lambda_c^{\rm pre}$ and invasion threshold $\lambda_c^{\rm inv}$,
   respectively.  Red circles (blue up triangles) are numerical
   simulations with $\rho_0=0.4$ ($0.07$). }
 \label{fig2}
\end{center}
\end{figure}

Figure~\ref{fig2} shows the social contagions in the community
networks. We find that the final behavior adoption size $\rho_a$ in
community $a$ increases discontinuously with the information
transmission probability $\lambda$, i.e., there is a first-order phase
transition that depends on $\rho_0$ and $\lambda$. For a small value of
the initially adopted seeds $\rho_0=0.07$, $\rho_a$ increases
discontinuously at the presence threshold $\lambda_c^{\rm pre}$, i.e.,
there is a vanishingly small fraction of individuals adopting the
behavior when $\lambda\leq\lambda_c^{\rm pre}$, and a finite fraction of
individuals adopting the behavior when $\lambda>\lambda_c^{\rm pre}$.

We find a similar phenomenon for a large seed size $\rho_0=0.4$, i.e.,
$\rho_b$ increases discontinuously with $\lambda$ at the invasion
threshold $\lambda_c^{\rm inv}$. These phenomena indicate that the
system exhibits first-order phase transitions with a hysteresis
loop. Specifically, the fraction of adopted individuals versus $\lambda$
depends on the initial conditions of $\rho_0$ at region $\lambda_c^{\rm
  inv}<\lambda< \lambda_c^{\rm pre}$. In this region, for a small
fraction of seeds, i.e., $\rho_0$ = 0.07, susceptible individuals from
both communities are less likely to receive a number of information
units that exceeds the adoption threshold. Large values of transmission
probability $\lambda$ are needed to accelerate social contagion. When
there is a large fraction of initial adopters, i.e., $\rho_0=0.4$, the
probability that the number of information units received by a
susceptible individual exceeds the adoption threshold increases. When
the values of the transmission probability $\lambda$ are small, the
contagion accelerates. The strength of the community structures does not
qualitatively affect the phenomena. Figure~2 shows that our theoretical
results agree with the numerical simulation results.

\begin{figure}
\begin{center}
 \includegraphics[scale=0.6]{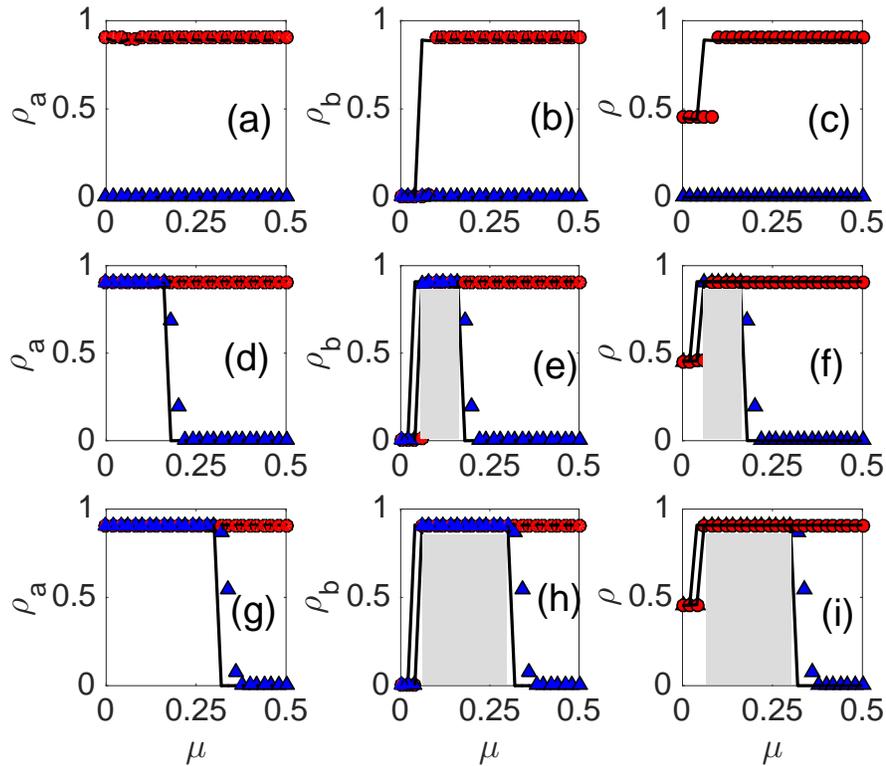}
 \caption{(Color online) Effects of the strength of community structures
   on social contagions. The final adoption size versus $\mu$ with
   $\lambda=0.5$ in Figs.~(a)--(c), $\lambda=0.7$ in Figs.~(d)--(f),
   $\lambda=0.9$ in Figs.~(g)--(i).  The three columns respectively
   represents the final behavior adoption size in community $a$, $b$ and
   the system. Red circles (blue up triangles) are numerical simulations
   with $\rho_0=0.4$ ($0.07$).  The lines are the theoretical
   predictions. The gray areas in Figs. (e), (f), (h) and (i) represent
   the optimal community structure that diffuses in global network. }
 \label{fig3}
 \end{center}
\end{figure}

We next determine the effect of community structure $\mu$ under
differing initial conditions (see Fig.~\ref{fig3}). As in
Fig.~\ref{fig2}, we find a hysteresis loop phenomenon, i.e., $\rho$
($\rho_a$ or $\rho_b$) may have different values under different initial
seed sizes. In community $a$, irrespective of the proportion of
intercommunity links ($\mu$), the internal connectivity can spread the
contagion to the entire originating community $a$ when $\rho_0$ is large
($\rho_0=0.4$), as shown in Figs.~\ref{fig3}(a), \ref{fig3}(d), and
\ref{fig3}(g). Figures~\ref{fig3}(d) and \ref{fig3}(g) show that
increasing $\lambda$, i.e., $\lambda=0.7$ and $\lambda=0.9$, when $\mu$
is small activates the modular structure in the originating community by
a small $\rho_0$ value. As $\mu$ increases, more intralinks (within
communities) are replaced by interlinks (between two communities). When
$\mu$ is large, individuals in community $a$ are less likely to expose
adopted neighbors. When $\mu$ is increased, the number of susceptible
individuals adopting the information in community $a$
decreases. Although susceptible individuals in community $b$ acquire
more adopted neighbors in community $a$, their number does not exceed
$\theta$.  Individuals in community $b$ have no adopted
state. Increasing $\mu$ prevents the contagion from spreading to the
entire network through internal connectivity. In community $b$ when both
$\rho_0$ and $\mu$ are small there are insufficient intercommunity
bridges to propagate social contagion from community $a$ to community
$b$, even when community $a$ is fully saturated [see Figs.~\ref{fig3}(e)
  and \ref{fig3}(h)]. Thus susceptible individuals in community $b$ have
too few adopted neighbors in community $a$ to receive information
sufficient to exceed the adoption threshold.

Figures~\ref{fig3}(e) and \ref{fig3}(h) show that increasing $\mu$
provides the optimal community structure for social contagions. Here the
system modularity is sufficiently large to initiate local spreading,
sufficiently small to induce intercommunity spreading, and the modular
structure allows intercommunity spreading from community $a$ to
community $b$. Thus social contagions exist in both communities $a$ and
$b$ in this region. If $\mu$ is too large, however, although there are
sufficient intercommunity bridges, the system modularity is too small to
initiate intercommunity spreading from community $a$. Because the
originating community is not saturated, the diffusion does not spread to
community $b$ [see Figs.~\ref{fig3}(d) and \ref{fig3}(g)]. When $\rho_0$
is large ($\rho_0=0.4$), the strong community structure enables
intercommunity spreading from the originating community $a$ to community
$b$. Again our theory agrees with the numerical simulations.

\begin{figure}
\begin{center}
 \includegraphics[scale=0.5]{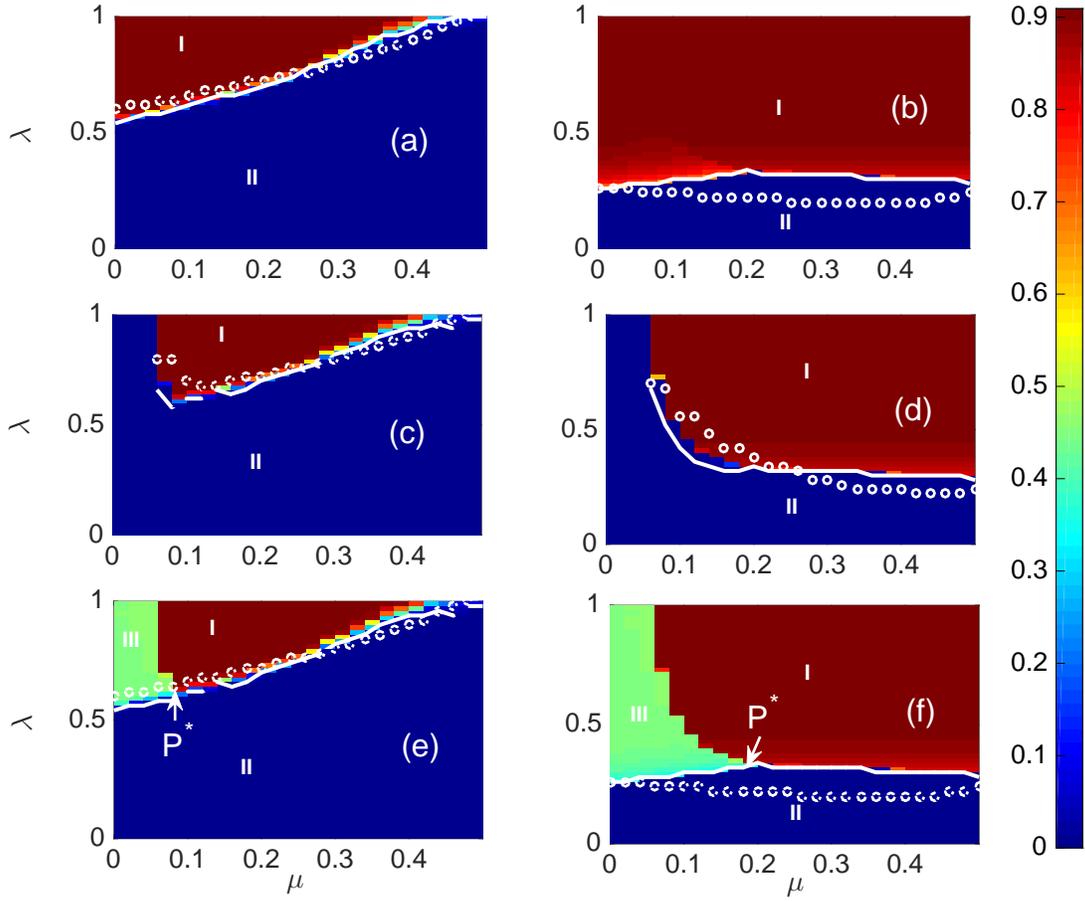}
\caption{(Color online) Phase diagram of the social contagions on plane
  $\mu-\lambda$. In (a), (c) and (e), we set $\rho_0=0.07$.  And in (b),
  (d) and (f), we set $\rho_0=0.4$. The symbols and lines are the
  numerical and theoretical predictions of the threshold points,
  respectively. The lines in (a)-(b), (c)-(d) and (e)-(f) represent
  $\rho_a$, $\rho_{b}$ and $\rho$, respectively. Region I (red), II
  (blue) and III (green) are global diffusion, no diffusion and local
  diffusion region, respectively. }
 \label{fig4}
 \end{center}
\end{figure}

Figure~\ref{fig4} shows the effects of $\lambda$ and $\mu$. Depending on
the fraction of the final behavior adoption size, the plane is divided
into phase diagrams: global diffusion (region I), no diffusion (region II),
and local diffusion (region III). The behavior of $\rho_{\infty}$ as a
function of $\mu$ and $\lambda$ exhibits qualitatively different
patterns depending on $\rho_0$. 

When $\mu$ is small, intralinks greatly outnumber interlinks. In
response to initially adopted seeds in community $a$, susceptible
community $a$ individuals are more likely to become adopted if the
number of received information units exceeds threshold $\theta$. When
there are fewer interlinks, community $b$ individuals are less likely to
receive message units that exceed the threshold, and the social
contagion remains local (region III). Increasing $\mu$ enables
susceptible community $b$ individuals to receive more message units from
exposed adopted neighbors in community $a$. Global diffusion (region I)
emerges when the message units that individuals in community $b$ receive
exceed threshold $\theta$. When there are few initial adopter seeds, the
probability that susceptible individuals have adopter neighbors
decreases as the number of intralinks decreases. When the number of
adopter seeds is too small to transmit sufficient message units to both
communities $a$ and $b$, the no-diffusion area (region II) appears. When
the information transmission probability $\lambda$ is too small, the
message units received by susceptible individuals in both communities do
not exceed $\theta$ and no susceptible individuals adopt the
information.

Figure~\ref{fig4}(e) shows that when $\rho_0=0.07$ is small and
community strength is intermediate and finite, $\mu$ allows global
spreading. However when $\mu$ is large the number of intracommunity
links is too small to propagate spreading in the originating community
$a$ and thus cannot be transmitted over the entire system, but when
$\rho_0=0.4$ is large [see Fig.~\ref{fig4}(f)] and larger than the
critical value for transition in a system without communities,
increasing $\mu$ does not block local spreading, and global diffusion
occurs only through external links. We find a rich phase diagram in the
$\mu$--$\lambda$ plane with a triple point $P^{*}$. As $\mu$ decreases,
the first order transition line that separates global diffusion (region
I) from no diffusion (region II) forks into two branches and generates a
new local diffusion phase (region III). Around $P^{*}$ a small variform
percentage of the edges between the communities can induce an abrupt
change in the number of adopted individuals.

\section{Conclusions}

\noindent
In this paper we have studied the reinfection pattern that most previous
research has ignored. Using infection thresholds we systematically
investigate how reinfection affects the social contagion dynamics in
community networks. We use a mean-field approximation approach that
produces results that agree with numerical simulation results.  We find
that first-order phase transitions exist during the spreading process in
communities, and that a hysteresis loop emerges when the spreading
probability at region $\lambda_c^{\rm inv}<\lambda<\lambda_c^{\rm pre}$
is in the system for different initial adopter densities. We also find
an optimal level of community structure strength that facilitates the
global diffusion of a small number of initially adopted seeds. In this
optimal community structure, global diffusion requires a minimal number
of adopters in the community. When the number of links between the
communities is decreased, we find a rich phase diagram with a triple
point.  Our numerical results agree with our proposed mean-field
approach, which quantifies, using threshold models, the influence of
reinfection in communal networks.

Our results use the initially adopted seeds in only one community. Using
numerical simulations and theoretical analyses, we find that our
conclusions are not qualitatively affected when the seeds are randomly
selected in two communities, and our theory produces results that agree
with simulation results when community networks are scale-free. In
addition, the amount of heterogeneity in the communal degree
distribution does not qualitatively affect these phenomena. Our findings
enrich our understanding of how social contagions transmit through
communal systems. Our theory in this work can be used to study epidemic
spreading \cite{Manlio,daqing, wangzhenplr,wangzhenchoas}, the effects
of vaccination \cite{wangzhenpr}, and the impact of human behavior
\cite{li,wangzhensa} on epidemics. In future work we will further
explore our approach using real social contagion data.

\section*{Acknowledgments}

\noindent
This work was funded in part by the National Key Research and
Development Program of China (Grant No. 2016YFB0800602), the Program for
Innovation Team Building of Mobile Internet and Big Data at Institutions
of Higher Education in Chongqing (Grant No. CXTDX201601021) and the
National Natural Science the Foundation of China (Grant
No. 61751110). The Boston University work was supported by DTRA Grant
HDTRA1-14-1-0017, by DOE Contract DE-AC07- 05Id14517, and by NSF Grants
CMMI 1125290, PHY 1505000, and CHE-1213217, and the LAB knowledge the
support of UNMdP and FONCyT, PICT 0429/13.

\end{document}